\documentclass[conference]{IEEEtran}
\IEEEoverridecommandlockouts

\newcommand\hmmax{0}
\newcommand\bmmax{0}

\usepackage{lipsum}
\usepackage{graphicx}
\usepackage{tikz}
\usepackage{pgfplots}
\usepackage{xcolor}
\usepackage{color, colortbl}
\usepackage{amsmath}
\usepackage{bbm}
\usepackage{amsfonts, amssymb}
\usepackage{mathtools}
\usepackage{cite}
\usepackage[nolist]{acronym}

\usepackage{booktabs} 		
\usepackage{diagbox}		
\usepackage{upgreek}
\usepackage{bbm}
\usepackage{Files/bm}
\usepackage{Files/winsnotation}
\usepackage{Files/Symbols_OPL}
\usepackage{Files/LVcolours} 

\usepackage[ruled,vlined,noresetcount]{algorithm2e}
\usepackage{setspace}	 	
\usepackage{comment}

\usepackage{array}
\newcolumntype{C}[1]{>{\centering\arraybackslash}p{#1}}

\makeatletter
\def\endthebibliography{%
  \def\@noitemerr{\@latex@warning{Empty `thebibliography' environment}}%
  \endlist
}
\makeatother
\IEEEoverridecommandlockouts

\usepackage{amsthm}
\theoremstyle{plain}

\theoremstyle{definition}

\newtheorem*{remark}{Remark}

\pgfplotsset{compat=1.17}

\begin{document}

\title{A Joint PHY and MAC Layer Design for \\ Coded Random Access with Massive MIMO}

\author{%
  \IEEEauthorblockN{Lorenzo Valentini, Marco Chiani, Enrico Paolini}
  \IEEEauthorblockA{CNIT/WiLab, DEI, University of Bologna, Italy\\
  	Email: \{lorenzo.valentini13, marco.chiani, e.paolini\}@unibo.it }
}

\maketitle

\begin{acronym}
\small
\acro{ACK}{acknowledgement}
\acro{AWGN}{additive white Gaussian noise}
\acro{BCH}{Bose–Chaudhuri–Hocquenghem}
\acro{BN}{burst node}
\acro{BS}{base station}
\acro{CDF}{cumulative distribution function}
\acro{CRA}{coded random access}
\acro{CRC}{cyclic redundancy check}
\acro{CRDSA}{contention resolution diversity slotted ALOHA}
\acro{CSA}{coded slotted ALOHA}
\acro{eMBB}{enhanced mobile broad-band}
\acro{FER}{frame error rate}
\acro{IFSC}{intra-frame spatial coupling}
\acro{i.i.d.}{independent and identically distributed}
\acro{IRSA}{irregular repetition slotted ALOHA}
\acro{LDPC}{low-density parity-check}
\acro{LOS}{line of sight}
\acro{MAC}{medium access control}
\acro{MIMO}{multiple input multiple output}
\acro{ML}{maximum likelihood}
\acro{MMA}{massive multiple access}
\acro{mMTC}{massive machine-type communication}
\acro{MPR}{multi-packet reception}
\acro{MRC}{maximal ratio combining}
\acro{PAB}{payload aided based}
\acro{PDF}{probability density function}
\acro{PGF}{probability generating function}
\acro{PHY}{physical}
\acro{PLR}{packet loss rate}
\acro{PMF}{probability mass function}
\acro{PRCE}{perfect replica channel estimation}
\acro{QPSK}{quadrature phase-shift keying}
\acro{RF}{radio-frequency}
\acro{SC}{spatial coupling}
\acro{SIC}{successive interference cancellation}
\acro{SIS}{successive interference subtraction}
\acro{SN}{sum node}
\acro{SNB}{squared norm based}
\acro{SNR}{signal-to-noise ratio}
\acro{URLLC}{ultra-reliable and low-latency communication}
\end{acronym}
\setcounter{page}{1}

\begin{abstract}
Grant-free access schemes are candidates to support future massive multiple access applications owing to their capability to reduce control signaling and latency.
As a promising class of grant-free schemes, coded random access schemes can achieve high reliabilities also with uncoordinated transmissions and therefore in presence packet collisions.
In this paper, an analysis tool for coded random access, based on density evolution, is proposed and exploited for system design and optimization.
In sharp contrast with the existing literature, where such tools have been developed under simplified channel assumptions, the proposed tool captures not only MAC layer features, but also the physical wireless fading channel and a realistic physical layer signal processing based on multiple antennas and randomly-chosen orthogonal pilots.
Theoretical results are validated by comparison with symbol-level Monte Carlo simulations.
\end{abstract}

\section{Introduction}

The ever-growing spread and pervasiveness of machine-type communications and of the Internet of Things has recently boosted the attention towards \ac{MMA} problems, where a massive number of devices, each with a sporadic but unpredictable activity, transmit short packets to a common \ac{BS} \cite{Liu2018:massivePt1,sanguinetti2018:random,Sor2018:coded}. 
A main target for \ac{MMA} protocols is to achieve a very high scalability, defined as the number of simultaneously active users that the system can support, in presence of latency and reliability constraints \cite{Hasan2013:random,Liu2018:sparse,Chen2020:massive}. 
In this context, grant-free multiple access schemes (as opposed to grant-based ones) can drastically reduce control signalling for connection establishment, bringing benefits in terms of both scalability and latency. 

A class of grant-free \acf{MAC} schemes that has attracted an increasing interest in the past few years, both in satellite and terrestrial applications, is that of \ac{CRA} schemes \cite{paolini2015:Coded,Berioli2016:Modern}. 
Based on a very simple uncoordinated access mechanism on the device side and on \ac{SIC} across different slots on the \ac{BS} side, they are intrinsically related to iterative decoding of codes on sparse graphs.
Instances of \ac{CRA} schemes include \ac{CRDSA} \cite{casini2007:contention}, \ac{IRSA} \cite{liva2011:irsa}, and \ac{CSA} \cite{paolini2015:csa}.

While several works on \ac{CRA} have addressed the collision channel model owing to its amenability to analysis, more recently the focus has been shifted to wireless channels models with realistic \acf{PHY} layer processing.
In this respect, an important direction of investigation focuses on \ac{CRA} in contexts where the receiver is a \ac{BS} with a massive number of antennas (e.g., \cite{DeCarvalho2017:Random,Sor2018:coded,Valentini2021:Coded,Valentini2022:Impact}).
This interest is justified by the possibility to exploit massive \ac{MIMO} to implement forms of \ac{MPR}, meaning that multiple packets can be correctly decoded in a single slot with a consequent substantial boost in scalability.

To achieve a high scalability under tightening reliability constraints, \ac{CRA} schemes necessitate of a careful design.
For example, with reference to the \ac{IRSA} scheme, this typically means finding an optimum probability distribution for the packet repetition rate employed by the generic active user.
The solution to this problem benefits from the bridge with codes on sparse graphs, so that it is usually tackled by tools developed in coding theory, like density evolution \cite{Richardson2001:design}.
So far, the design of \ac{CRA} schemes is well-established over channel models, such as the collision channel \cite{liva2011:irsa,paolini2015:csa} or its \ac{MPR} extension \cite{stefanovic2018:multipacket}, where the communication process over the \ac{PHY} channel as well as signal processing at \ac{PHY} layer are modeled in a very simple, often idealized, way. 

In this paper we aim at addressing this problem by extending the density evolution analysis of \ac{IRSA} to a more realistic setting.
More specifically, we consider \ac{IRSA} over a Rayleigh block fading channel with a massive \ac{MIMO} \ac{BS} and power control and we show that it is possible to tailor density evolution analysis, and consequent system optimization, to this channel model and to a specific \ac{PHY} signal processing based on randomly-chosen orthogonal pilots for channel estimation and \ac{MRC}. 
Notably, the results are easily extendable to the more general \ac{CSA} setting.
The main contributions of the paper can be summarized as follows:
\begin{itemize}
    \item We extend the simple collision channel to the case where orthogonal resources are available within each slot.
    \item We generalize \ac{CRA} density evolution analysis to capture both \ac{PHY} processing and Rayleigh fading.
    \item We show that \ac{IRSA} distributions that are optimum over the collision channel may turn sub-optimum when a realistic setting is considered.
\end{itemize}

The paper is organized as follows. Section~\ref{sec:preliminary} introduces preliminary concepts, the system model, and some background material. 
Section~\ref{sec:DEAdHocCh} addresses density evolution generalizations. 
Numerical results are provided in Section~\ref{sec:NumericalResults}. Finally, conclusions are drawn in Section~\ref{sec:Conclusions}.

\section{Preliminaries and Background}
\label{sec:preliminary}

We assume that the time is organized in \ac{MAC} frames, each composed by $N_\mathrm{s}$ slots. 
In a frame time, $K_\mathrm{a}$ active users (with $K_{\mathrm{a}}$ random and unknown to the receiver) contend for transmission of one information packet each.
Considering an \ac{IRSA} access protocol, the generic user sends $r$ replicas of its packet (or ``burts'') in $r$ randomly-chosen slots. For all users, $r$ is generated randomly according to the same probability distribution.
As remarked later, although we stick to \ac{IRSA}, the developed results have a broader applicability to \ac{CRA} schemes. 


\subsection{Density Evolution over the Collision Channel} \label{subsec:DECollisionCh}

This section reviews density evolution equations for \ac{IRSA} over the collision channel \cite{liva2011:irsa}.
The system can be represented by a bipartite graph where $K_\mathrm{a}$ user nodes, also known as \acp{BN}, are connected with $N_\mathrm{s}$ slot nodes, also referred to as \acp{SN}. 
A \ac{BN} corresponding to an active user sending $r$ packet replicas has $r$ edges towards the $r$ \acp{SN} associated with the chosen slots.
The repetition degree $r$ is a random variable with \ac{PGF} $\Lambda(x) = \sum_r \Lambda_r x^r$, independently drawn by each active user. 
Then, the probability that an edge is connected to a degree-$r$ \ac{BN}, $\lambda_r$, is given by
\begin{align}
    \lambda_r = \frac{\Lambda_r \, r}{\Lambda'(1)}\,.
\end{align}
On the other hand, a \ac{SN} has $c$ edges representing the number of users that have selected the corresponding slot to transmit a packet replica.
We denote by $\rho_c$ the probability that an edge is connected to a \ac{SN} of degree $c$; this is given by
\begin{align}
    \rho _c = \frac{\Psi_c \, c}{\sum_h \Psi_h \, h}
\end{align}
where $\Psi_c$ is probability that $c$ users performed a transmission in the considered slot.

The \ac{PGF} $\Lambda (x)$, or \ac{BN} degree distribution (or simply \ac{IRSA} distribution), is under the control of the system designer. 
This is indeed not the case for the \ac{SN} degree distribution $\Psi (x) = \sum_c \Psi_c \, x^c$, that is fully defined by the system load $G$ and by the average burst repetition rate, $\sum_r r \, \Lambda_r = \Lambda' (1)$.
In particular, for a large users' population size $K$, it is licit to assume that the number of transmissions in a slot follows a Poisson distribution.
Specifically, we can write
\begin{align}
    \Psi_c = \frac{(G\, \Lambda'(1))^c}{c!} \, \text{exp}(-G\, \Lambda'(1))
\end{align}
which also yields
\begin{align}\label{eq:rho_c}
    \rho_c = \frac{(G\, \Lambda'(1))^{c-1}}{(c-1)!} \, \text{exp}(-G\, \Lambda'(1)) .
\end{align}

The collision channel assumptions can be summarized as:
\emph{Assumption~1}: If the number of arrivals in a slot is larger than one, then the receiver is unable to successfully decode any of these bursts.
\emph{Assumption~2}: If there is only one arrival in a slot, then the burst is successfully decoded with zero error probability and perfectly subtracted from the slot.
\emph{Assumption~3}: Whenever a burst is successfully decoded in a slot, the interference generated by each of its replicas can be perfectly subtracted from the corresponding slots.

Under these assumptions, let $q_\ell^{(r)}$ be the probability that an edge, connected to a degree-$r$ \ac{BN}, is unknown at the end of \ac{SIC} iteration $\ell$. 
Let us also define $p_\ell^{(c)}$ as the probability that an edge, connected to a degree-$c$ \ac{SN}, is unknown at the end of \ac{SIC} iteration $\ell$. 
We can exploit the edge-oriented probabilities $\lambda_r$ and $\rho_c$ to define the average probabilities $q_{\ell}$ and $p_{\ell}$ as 
\begin{align}
\label{eq:qell_pell_General}
    q_\ell = \sum_r \lambda_r \, q_\ell^{(r)} \quad \mathrm{and} \quad p_\ell = \sum_c \rho_c \, p_\ell^{(c)}\,.
\end{align}

Next, consider a degree-$r$ \ac{BN}. 
An edge is revealed whenever any of the other edges connected to the same \ac{BN} has been revealed. 
Thus, the probability that a burst has not yet been retrieved by the ``\ac{MAC} layer repetition code'' is
\begin{align}
\label{eq:qell_d_IRSA}
    q_\ell^{(r)} = p_{\ell - 1}^{r - 1}\,.
\end{align}
Similarly, consider a degree-$c$ \ac{SN}. 
An edge is revealed whenever all the other edges have been revealed due to the collision channel assumptions. Hence, the probability that a burst in a slot has not yet been cancelled by \ac{SIC} is
\begin{align}
\label{eq:pell_c_Collision}
    p_\ell^{(c)} &= 1 - (1 - q_\ell)^{c - 1} \, .
\end{align} 
Combining equations from \eqref{eq:rho_c} to \eqref{eq:pell_c_Collision} leads to the density evolution recursion over the collision channel in the form
\begin{align}\label{eq:p_ell_Collision}
    p_\ell &= 1 - \exp \Big( -G\, \sum_r r \Lambda_r p_{\ell-1}^{r-1} \Big)
\end{align}
with initial condition $p_0 = 1$ (there are no revealed edges at the beginning of the process). The asymptotic \acl{PLR} at \ac{SIC} iteration $\ell$ is given by
\begin{align}
    Q_\ell =  \sum_r \Lambda_r p_\ell^r \,.
\end{align}
Finally, the asymptotic load threshold of an \ac{IRSA} distribution $\Lambda(x)$ is defined as
\begin{align}
\label{eq:Gstar}
    G^{*} = \sup \{ G > 0 : Q_\ell \rightarrow 0 \text{ as } \ell \rightarrow \infty\}
\end{align}
where we note that $Q_\ell \rightarrow 0$ if and only if $p_\ell \rightarrow 0$.
Here, ``asymptotic'' refers to the fact that density evolution recursion assumes statistical independence of messages along the edges of the graph, a condition that is met in the limit as $K \rightarrow \infty$, $N_\mathrm{s} \rightarrow \infty$, and the ratio $K / N_\mathrm{s}$ being constant.

\begin{remark}
We highlight that, in density evolution recursion of \ac{CRA} protocols, $q_\ell$ depends on the \ac{MAC} layer, while $p_\ell$ on the channel model and \ac{PHY} layer processing. 
For this reason, the results developed in this paper, aimed at evaluating $p_\ell$ over a realistic channel model and \ac{PHY} processing, are amenable to extension to other \ac{MAC} \ac{CRA} protocols such as \ac{CSA} \cite{paolini2015:csa}.
\end{remark}


\subsection{Scenario Definition}
\label{subsec:Scenario}

The availability of a \ac{BS} with a massive number of antennas is a key feature to enable \ac{MPR} (i.e., decoding of multiple packets in the same slot even without \ac{SIC} across slots) at the receiver. 
A simple approach to achieve \ac{MPR} is represented by use of orthogonal pilots in combination with massive \ac{MIMO} processing \cite{Sor2018:coded,Valentini2021:Coded,Valentini2022:Impact}.
Since in \ac{MMA} applications the number of users is typically extremely large compared with the number of available pilots, $N_\mathrm{P}$, it is however not possible to pre-assign one specific pilot to each user. 
Rather, each active user may pick one pilot randomly from the set of $N_\mathrm{P}$ orthogonal ones, without any coordination with the other users (hence, possibly yielding pilot collisions in a slot). 
In combination with \ac{IRSA}, this approach was dubbed ``coded pilot random access'' in \cite{Sor2018:coded}.

Next, we review the \ac{PHY} layer model and processing adopted in this paper, which is the one also considered in \cite{Valentini2021:Coded,Valentini2022:Impact}.
The model may be summarized as follows:
\begin{itemize}
    \item The channel is a Rayleigh block fading channel with power control and coherence time equal to the slot time.
    \item The receiver has $M$ antennas, each with independent fading coefficient per user.
    \item An active user picks, for each burst, a pilot uniformly at random from a set of $N_\mathrm{P}$ orthogonal ones for channel estimation purposes.
    \item \Ac{QPSK} modulation with Gray mapping and hard decision is employed.
    \item The payload is composed by $N_\mathrm{D}$ symbols and protected using a \ac{BCH} code able to correct up to $t$ errors per codeword.
    \item The receiver performs a ``channel hardening-based'' \ac{SIC} processing, as described in \cite{Sor2018:coded,Valentini2021:Coded,Valentini2022:Impact}.
\end{itemize}
Then, the received signal in a slot may be expressed as $[\M{P}, \M{Y}] \in \mathbb{C}^{M \times (N_\mathrm{P}+N_\mathrm{D})}$ where
\begin{align}\label{eq:P}
    \M{P} = \sum_{k \in \mathcal{A}} \V{h}_k \V{s}(k) + \M{Z}_p \,\,\,\, \mathrm{and} \,\,\,\, \M{Y} = \sum_{k \in \mathcal{A}} \V{h}_k \V{x}(k) + \M{Z} .
\end{align}
In \eqref{eq:P}, $\mathcal{A}$ is the set of users transmitting a burst in the considered slot; $\V{h}_k = (h_{k,1}, \dots, h_{k,M})^T \in \mathbb{C}^{M\times 1}$ is the $k$-th user channel coefficient vector, whose elements are \ac{i.i.d.} random variables with distribution $\mathcal{CN}(0, \sigma_\mathrm{h}^2)$ for all $k \in \mathcal{A}$ due to power control. Without losing generality it is assumed $\sigma_\mathrm{h}^2 = 1$.
Moreover, $\V{s}(k) \in \mathbb{C}^{1\times N_\mathrm{P}}$ and $\V{x}(k) \in \mathbb{C}^{1\times N_\mathrm{D}}$ are the pilot sequence and payload transmitted by user $k$.
Finally, $\M{Z}_p \in \mathbb{C}^{M\times N_\mathrm{P}}$ and $\M{Z} \in \mathbb{C}^{M\times N_\mathrm{D}}$ are matrices of Gaussian noise samples.

The processing is split into two phases. In phase~$1$, all slots are processed in order. 
In each slot, the \ac{BS} first attempts channel estimation for all possible pilots by computing $\V{\phi}_j\in \mathbb{C}^{M\times 1}$ and consequently attempting payload estimation, for all $j \in \{1,\dots,N_{\mathrm{P}}\}$, as
\begin{align}
\label{eq:PhiEstimate}
    \V{\phi}_j &= \frac{\M{P} \,\V{s}_j^{H}}{\| \V{s}_j \|^2} = \sum_{k \in \mathcal{A}^j} \V{h}_k + \V{z}_j\,\,\,\, \mathrm{and} \,\,\,\,  \hat{\V{x}}_j = \frac{\V{\phi}_j^{H} \, \M{Y}}{\| \V{\phi}_j \|^2}\,.
\end{align}
Here, $\mathcal{A}^j$ is the set of active users employing pilot $j$ in the current slot, $\V{s}_j \in \mathbb{C}^{1\times N_\mathrm{P}}$ is the $j$-th pilot sequence, and $\V{z}_j \in \mathbb{C}^{M \times 1}$ is a noise vector.

If a generic user $\ell$ is the only one picking pilot $j$ in the currently processed slot ($\mathcal{A}^j = \{ \ell \}$) and if the number of antennas $M$ is large enough, then $\hat{\V{x}}_j$ approximates the user's payload $\V{x}(\ell)$.
Upon successful channel decoding performed on $\hat{\V{x}}_j$, the packet symbols are stored in a buffer waiting for the \ac{SIC} phase. 

In phase~$2$, \ac{SIC} is performed across slots relying on information about replica number and positions along with the employed pilots for each of them retrieved in the decoded payload.
This can be implemented, for example, by letting the number of replicas, their positions, and their pilot indexes be functions of the random message.
By subtracting interference, it is possible that new users can be found re-attempting channel decoding procedure.
Then, whenever a new user is successfully decoded 
its information is stored and phase $2$ iterates until the buffer contain no more users. More details on the \ac{SIC} and channel model, here omitted for space reasons, are available in \cite{Sor2018:coded,Valentini2021:Coded, Valentini2022:Impact}.

\section{Density Evolution with Realistic PHY Layer} 
\label{sec:DEAdHocCh}

In this section we develop a density evolution analysis for \ac{IRSA} over the realistic channel model and \ac{PHY} processing that was reviewed in Section~\ref{subsec:Scenario}.
This analysis is presented in Section~\ref{subsec:RealisticChannel}.
Before that, in Section~\ref{subsec:ColChOverRes}, we extend the above-reviewed density evolution analysis over the collision channel to the case of a collision channel featuring $N_{\mathrm{P}}$ orthogonal resources per slot.
To keep a clean and compact notation, we denote the probability that a random variable $\rv{A}$ takes the value $a$, $\Pr(\rv{A} = a)$, as $P(a)$. 
Similarly, we write $P(a, b|c)$ to indicate the probability $\Pr(\rv{A} = a, \rv{B} = b \,|\, \rv{C} = c)$, and $P(\mathcal{E})$ to denote the probability that an event $\mathcal{E}$ occurs.

\subsection{Collision Channel with Orthogonal Resources}\label{subsec:ColChOverRes}

Consider a simple extension of the collision channel in which $N_{\mathrm{P}}$ orthogonal resources are available in each slot of the frame.
In \ac{IRSA} over this channel, an active user generates its repetition rate $r$ according to the distribution $\Lambda(x)$, chooses $r$ slots uniformly at random (without replacement), then chooses one resource per slot uniformly at random, and finally transmits $r$ replicas of its packet into these slot-resource pairs.
For this channel, Assumptions~1-3 listed in Section~\ref{subsec:DECollisionCh} remain valid, upon simply replacing ``slot'' with ``slot-resource pair''. 
Note that the collision channel corresponds to $N_{\mathrm{P}}=1$.

To extend \ac{IRSA} density evolution equations to this channel, we initially observe that $\rho_c$ in \eqref{eq:rho_c}, $q_\ell$ in \eqref{eq:qell_pell_General}, and $q_\ell^{(r)}$ in \eqref{eq:qell_d_IRSA} remain unchanged as they only depend on the \ac{IRSA} multiple access protocol.
To update the expression of $p_{\ell}^{(c)}$, consider a slot with $c$ burst arrivals. 
Consider any such burst and let $\rv{H}$, $0 \leq \rv{H} \leq c-1$, be a random variable representing the number of interferers not yet recovered in the slot at the current iteration $\ell$. 
Defining the failure event $\mathcal{F}_\ell = \{$The replica of an edge is not yet decoded at iteration~$\ell \}$, we have
\begin{align}
\label{eq:plcDerivationOverRes}
p_\ell^{(c)} &= P(\mathcal{F}_\ell | c) = \sum_{h=0}^{c-1} P(h) \, P(\mathcal{F}_\ell | c, h ) \notag \\
&= \sum_{h=0}^{c-1} \binom{c-1}{h}\, q_\ell^h \,(1-q_\ell)^{c-1-h} \Big[ 1 - \Big(1 - \frac{1}{N_{\mathrm{P}}} \Big)^h \Big] \notag \\
&= 1 - \Big( 1 - \frac{q_\ell}{N_{\mathrm{P}}} \Big)^{c-1} \, .
\end{align}
where the factor $[ 1 - (1 - 1/N_{\mathrm{P}})^h ]$ is the probability that at least one of the $h$ interferers collides with the considered burst in the same resource.
Repeating the steps reviewed for the collision channel, we obtain density evolution recursion
\begin{align}\label{eq:p_ell_Collision_resources}
    p_\ell = 1 - \exp \Big( - \frac{G}{N_{\mathrm{P}}}\, \sum_r r  \Lambda_r \, p_{\ell-1}^{r-1} \Big)\,.
\end{align}
From \eqref{eq:p_ell_Collision_resources} and \eqref{eq:p_ell_Collision} we deduce that, over the collision channel with $N_{\mathrm{P}}$ orthogonal resources per slot: (i) the asymptotic threshold $G^*$ of a distribution $\Lambda(x)$ is $N_{\mathrm{P}}$ times larger than the threshold over the collision channel for the same distribution; (ii) for a given average number of replicas $\Lambda'(1)$, the distribution that maximizes the threshold $G^*$ over the collision channel remains optimum also on the collision channel with $N_{\mathrm{P}}$ orthogonal resources per slot.

\subsection{MIMO Block Fading Channel with Realistic Processing}\label{subsec:RealisticChannel}
In this section we extend density evolution equations to the scenario described in Section~\ref{subsec:Scenario}. 
Referring to the collision channel assumptions, pointed out in Section~\ref{subsec:DECollisionCh} and generalized in Section~\ref{subsec:ColChOverRes} to $N_{\mathrm{P}} > 1$, we note that Assumption~1 still holds due to the hypothesis of power control. 
On the other hand, Assumptions~2 and 3 do not hold any more because of the realistic payload and channel estimations. 
As the \ac{PHY} processing and interference cancellation in a slot are captured by \eqref{eq:pell_c_Collision}, next we focus on $p_\ell^{(c)}$ under realistic assumptions.

In density evolution, the probability $p_\ell^{(c)}$ is associated with the occurrence of the event that a packet replica, arriving in a slot where $c$ users have transmitted (i.e., with $c-1$ interfering bursts), is not successfully decoded in that slot at iteration $\ell$.
From a user's viewpoint, we can therefore define the random variable $\rv{I}$, representing the number of other interfering bursts that have picked  the same pilot (pilot-colliding bursts). 
The random variable $\rv{I}$ is considered to account for the behaviour explicitly shown in \cite{Valentini2022:Impact} in which the probability to fail the decoding of a user packet depends on the initial number of user choosing the same pilot.
We can write the probability that a user has exactly $i$ pilot-colliders, given $c$ total arrivals in the slot and $N_\mathrm{P}$ available pilots, as
\begin{align}
    P(i | c) = 
    \binom{c-1}{i} \, \left(\frac{1}{N_\mathrm{P}}\right)^i \, \left(1 - \frac{1}{N_\mathrm{P}}\right)^{c-1-i}\,.
\end{align}
Since to successfully decode a burst within a slot it is necessary that it is the only one with a given pilot, we define the random variable $\rv{S}$ as the number of pilot-colliding bursts that have been subtracted so far. Then, considering that from previous interference cancellations bursts can be subtracted with probability $1 - q_{\ell}$, the probability that exactly $s$ subtractions have been performed, given $i$ pilot-colliding users, is
\begin{align}
    P(s | i) = 
    \binom{i}{s} \, \left(1 - q_{\ell}\right)^s \, q_{\ell}^{i-s}\,.
\end{align}
Noting that $P(s | i, c) = P(s | i)$, we can write
\begin{align}
\label{eq:plcDerivation}
    p_\ell^{(c)} &= \sum_i \sum_s P(\mathcal{F}_\ell| i, s, c) \, P(s | i)  \, P(i | c) \,.
\end{align}
Since it is not possible to successfully decode a pilot-collided replica due to the power control hypothesis, we have $P(\mathcal{F}_\ell| i, s, c) = 1$, for all $s \neq i$.
On the other hand, when $s = i$ and a realistic channel is considered, there is a non-zero probability to successfully decode the burst. 
Using the approximation analytically derived in \cite{Valentini2022:Impact}, we can write
\begin{align}
    \label{eq:PFisc}
    P(\mathcal{F}_\ell| i, s, c) = \begin{dcases} 
    P_\mathrm{fail}((i+1) c - 1),
    & s=i\\
    1, & s \neq i\\
    \end{dcases} 
\end{align}
where
\begin{align}
    \label{eq:Pfail1}
    P_\mathrm{fail}(n) &= 1 - \sum_{d = 0}^{t} \binom{N_\mathrm{D}}{d}\, P_\mathrm{e}^d(n) \left( 1 - P_\mathrm{e}(n) \right)^{N_\mathrm{D}-d}
\end{align}
and
\begin{align}
    \label{eq:Pfail2}
    P_\mathrm{e}(n) &= \text{erfc}\left(\sqrt{\frac{M}{2 n}}\right) - \frac{1}{4} \text{erfc}^2\left(\sqrt{\frac{M}{2 n}}\right) \, .
\end{align}
It is possible to note that $P_\mathrm{fail}$ depends on the number of payload symbols $N_\mathrm{D}$, the number of antennas $M$, the error correction capability $t$ of the \ac{PHY} channel code, and the number of interfering users. 
Finally, we can write
\begin{align}\label{eq:plc}
    p_\ell^{(c)} 
    &= 1 - \left( 1 - \frac{q_{\ell}}{N_\mathrm{P}} \right)^{c-1} \notag \\ &+ \sum_{i=0}^{c-1} \left(1 - q_{\ell}\right)^i P_\mathrm{fail}((i+1) c - 1) \, P(i | c) \, .
\end{align}
Using $p_\ell^{(c)}$ from \eqref{eq:plc} instead of \eqref{eq:pell_c_Collision} in the density evolution recursion, allows us evaluating the asymptotic threshold $G^{*}$ of an \ac{IRSA} distribution $\Lambda(x)$ over the realistic wireless channel and signal processing.

\begin{remark}
Equation \eqref{eq:PFisc} accounts for the fact that, in realistic settings, the above collision channel Assumptions~2 and 3 do not hold.
In this respect we note that, letting $P(\mathcal{F}_\ell| i, s, c) = 1- \delta_{i,s}$ where $\delta_{i,s}$ is the discrete delta function, it is easy to retrieve \eqref{eq:plcDerivationOverRes}. This observation confirms that the procedure derived in this section can be seen as a ``generalization'' of the density evolution over collision channel.
\end{remark}

\section{Performance Evaluation}
\label{sec:NumericalResults}

As main outcomes of our analysis: (i) we use the developed analysis tool with differential evolution optimization \cite{storn1997differential} to derive optimal $\Lambda(x)$ distributions; (ii) we verify the consistency between the optimization results and Monte Carlo simulations (with both \ac{PHY} and \ac{MAC} layers); (iii) we show that an over-simplification of the \ac{PHY} channel model may lead to a wrong design when adopted in a realistic system. 
Before presenting the results, we briefly review differential evolution and the \ac{PHY} layer aware system adopted to verify the theoretical results, respectively in Section~\ref{subsec:DiffEvo} and Section~\ref{subsec:SimSetUp}. 

\subsection{Differential Evolution Optimization}
\label{subsec:DiffEvo}

We search for the optimal $\Lambda(x)$ distribution which maximizes $G^{*}$ in \eqref{eq:Gstar}.
The optimization problem is constrained by imposing the average number of transmitted packets $\Lambda^{\prime}(1)$, related to the average energy spent per user packet.
We solved the optimization problem using differential evolution \cite{storn1997differential}, an evolutionary, population-based metaheuristic search algorithm used to find the global minimum of a real-valued function of a vector of continuous parameters. 
The main steps are similar to those of evolutionary optimization algorithms \cite{back1997handbook}.
An initial population of vectors, each representing a $\Lambda(x)$ distribution, is first generated. 
A competitor for each population element is then constructed by mutation and crossover over the current vector population. 
Next, each population element is compared with its own competitor and only one is selected, resulting in an evolved population. 
The mutation, crossover and selection steps are iterated until a certain stopping criterion is fulfilled.
Differential evolution was proposed for the optimization of \ac{LDPC} codes degree profile in \cite{shokrollahi2005design} and for \ac{IRSA} and \ac{CSA} optimization in \cite{liva2011:irsa} and \cite{paolini2015:csa}.


\subsection{Simulation Setup}
\label{subsec:SimSetUp}
With reference to Section~\ref{subsec:Scenario}, we consider a system where users transmit payloads encoded with an $(n = 511, k = 421, t = 10)$ binary \ac{BCH} code. Part of the information bits are used to validate the decoded packets, for example by \ac{CRC}.
After zero padding the \ac{BCH} codeword with a final bit, we map the encoded bits onto a \ac{QPSK} constellation with Gray mapping, obtaining $N_\mathrm{D}=256$ symbols per codeword.
Simulations have been carried out with symbol rate $B_\mathrm{s} = 1$~Msps and $M = 256$ \ac{BS} antennas.
Imposing a maximum latency constraint of $\Omega = 50$~ms, we compute the number of slots per frame $N_\mathrm{s}$ as \cite{Valentini2021:Coded}
\begin{align}
\label{eq:NSlot}
    N_\mathrm{s} = \left\lfloor \frac{\Omega \, B_\mathrm{s}}{2 \, (N_\mathrm{P} + N_\mathrm{D})} \right\rfloor
\end{align}
where the number of orthogonal pilot symbols, $N_\mathrm{P}$, equals the total number of available pilot sequences.
These sequences are constructed using Hadamard matrices. 
In particular, we assume $N_\mathrm{P} = 64$, resulting in a number of slots $N_\mathrm{s} = 78$.
\subsection{Numerical Results}
\label{subsec:NumRes}

\begin{figure}[t]
    \centering
    \includegraphics[width=0.99\columnwidth]{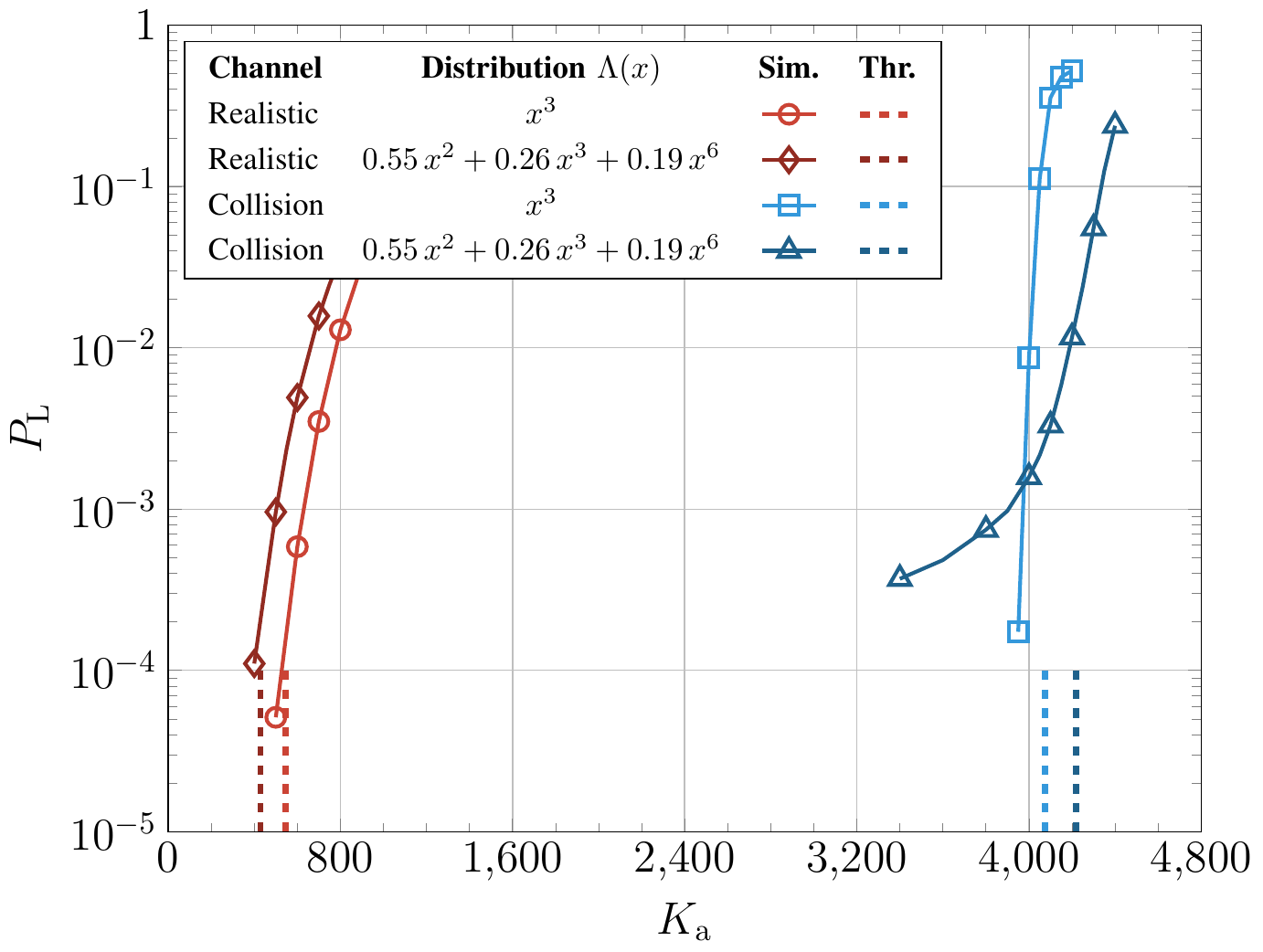}
    \caption{Packet loss rate comparison between the  \ac{IRSA} distribution with $\Lambda'(1) = 3$ and maximum repetition degree $6$ being optimal over the collision channel (with or without orthogonal resources) and the concentrated distribution $\Lambda(x)=x^3$.  
    Channels: Collision channel with $N_{\mathrm{P}}$ orthogonal resources and \ac{MIMO} block fading  channel with realistic signal processing. 
    Parameters: $N_\mathrm{P} = 64$, $N_\mathrm{s} = 78$, $M = 256$. 
    Dashed lines: Values of $G^* N_{\mathrm{s}}$. 
    }
    \label{fig:IRSAvsCRDSA_mean3}
\end{figure}

\begin{figure}[t]
    \centering
    \includegraphics[width=0.99\columnwidth]{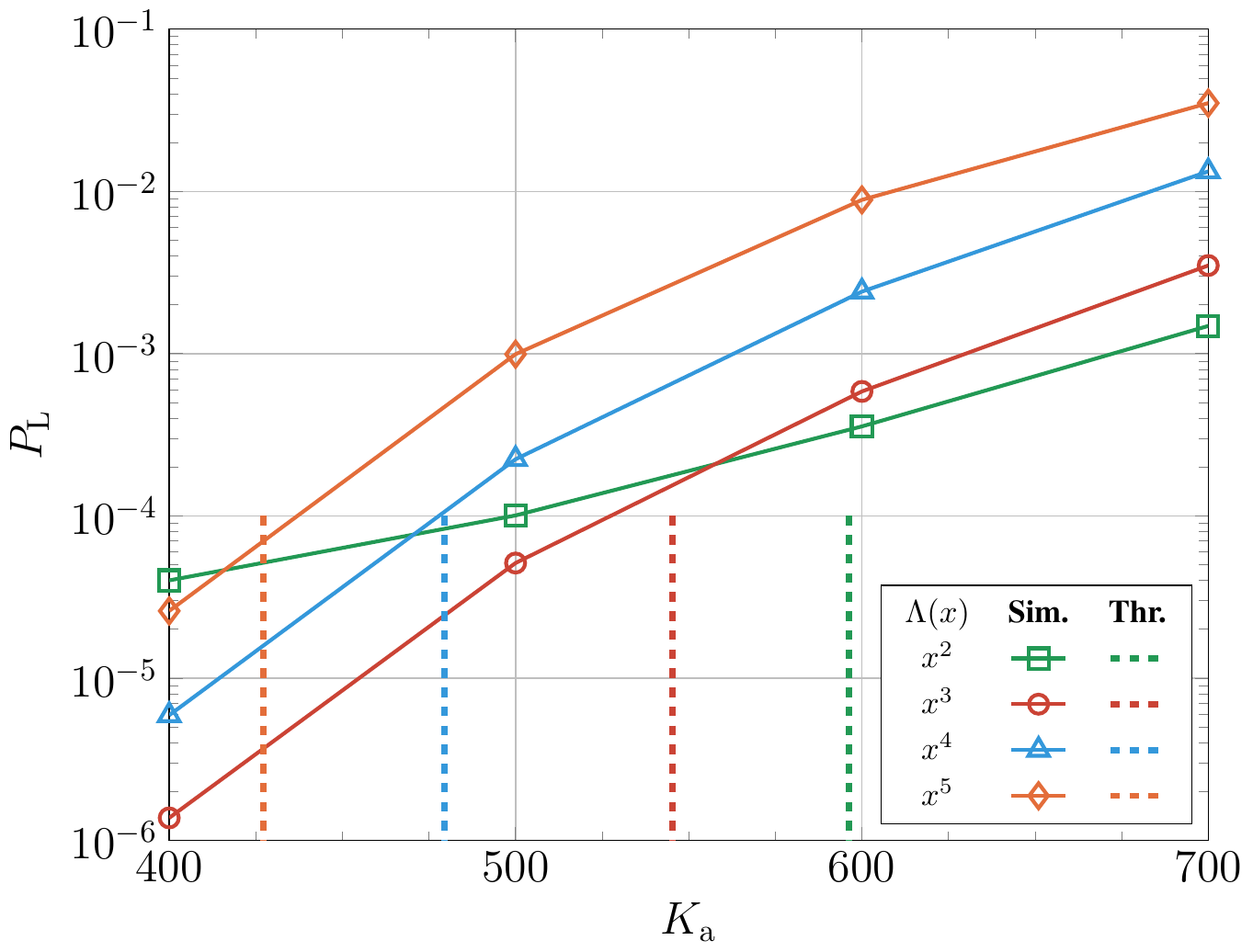}
    \caption{Packet loss rate  comparison between concentrated distributions characterized by different repetition rates $r \in \{ 2, 3, 4, 5\}$ over realistic channel. Solid: Monte Carlo simulations.  
    Parameters: $N_\mathrm{P} = 64$, $N_\mathrm{s} = 78$, and $M = 256$. Dashed lines: $N_\mathrm{s} G^*$.}
    \label{fig:CRDSAthrDensity}
\end{figure}

We start by presenting numerical results that illustrate the accuracy of the proposed threshold analysis. 
To this aim, we ran Monte Carlo simulations for some \ac{IRSA} distributions $\Lambda(x)$ over both the collision channel with orthogonal resources and the \ac{MIMO} block fading channel with actual signal processing, and performed threshold analysis for the same distributions over these channels.
In practice, we declared a value of $G$ as achievable (i.e., $G < G^{*}$) when density evolution recursion yielded $Q_\ell < 10^{-4}$ after a sufficiently large number of iterations.

In Fig.~\ref{fig:IRSAvsCRDSA_mean3} and Fig.~\ref{fig:CRDSAthrDensity} we report simulation results (in terms of \acl{PLR} versus the number of active users over the frame) in solid lines, while thresholds are marked by dashed vertical lines. 
In these figures, the ``thresholds'' are defined as $K_\mathrm{a}^{*} = N_\mathrm{s} G^{*}$, which represents an approximation of the number of simultaneously active users the scheme can support.
Fig.~\ref{fig:IRSAvsCRDSA_mean3} shows that the \ac{IRSA} distribution with average packet repetition rate $\Lambda'(1) = 3$ and maximum repetition rate $6$, having the largest threshold over the collision channel model \cite{paolini2015:csa}, becomes sub-optimal when the realistic channel and signal processing is considered. 
In fact, its threshold is outperformed by that of the distribution with a constant repetition rate $\Lambda(x) = x^3$ (that is, \ac{CRDSA} with repetition rate $3$). 
Very remarkably, as predicted by our threshold analysis, this result is in perfect agreement with the Monte Carlo simulation.
Fig.~\ref{fig:CRDSAthrDensity} shows similar results for other  distributions, which again reveal the effectiveness and reliability of the proposed analysis over massive \ac{MIMO} block fading channels and realistic \ac{PHY} layer processing. 
Note that all concentrated (\ac{CRDSA}) distributions considered in Fig.~\ref{fig:CRDSAthrDensity} exhibit the best threshold constrained to the corresponding integer $\Lambda'(1)$. 
Looking again at Fig.~\ref{fig:CRDSAthrDensity} we can see that, as expected, the proposed density evolution analysis is unable to capture error floor phenomena such as the one affecting the distribution $\Lambda(x) = x^2$. 
In Table~\ref{tab:gStar} we list the thresholds estimated through density evolution for some $\Lambda(x)$ distributions with the previous choice of the system parameters.

Lastly, we show in Table~\ref{tab:VarM} the results of another analysis we carried out using the proposed tool. 
For a constrained average repetition rate $\Lambda'(1) = 3$, we let the number of \ac{BS} antennas $M$ vary, searching for the optimum distribution (in terms of $G^*$) for each considered $M$. 
For all values of $M$, differential evolution optimization returned the same distribution $\Lambda(x) = x^3$.
It is interesting to observe that, while the asymptotic threshold $G^*$ increases monotonically with $M$, the ratio $G^*/M$ (which represents a sort of efficiency per antenna) is not monotonically increasing but exhibits a maximum value.
We attribute the decrease of $G^*/M$ for large $M$ to the constant number of orthogonal pilots $N_{\mathrm{P}} = 64$.

\begin{table}[t]
    \centering
    \setlength{\tabcolsep}{10pt}
    \caption{Asymptotic thresholds obtained through density evolution under realistic channel assumptions.}
    \label{tab:gStar}
    \small
    \begin{tabular}{l | c c}
        \toprule
        \rowcolor[gray]{.95}
        IRSA Distribution & $\Lambda'(1)$ & $G^{*}$\\
        \midrule
        $\Lambda(x) = x^2$ & 2 & $7.64$ \\
        $\Lambda(x) = x^3$ & 3 & $6.99$ \\
        $\Lambda(x) = x^4$ & 4 & $6.15$ \\
        $\Lambda(x) = x^5$ & 5 & $5.48$ \\
        $\Lambda(x) = 0.55 \, x^2 + 0.26\, x^3 + 0.19\, x^6$ & 3 & $5.49$ \\
        $\Lambda(x) = 0.50 \, x^2 + 0.50 \, x^3$ & 2.5 & $6.64$ \\
        $\Lambda (x) = 0.51\, x^2 + 0.27\, x^3 + 0.22 \, x^8$ & 3.6 & $4.63$ \\
        $\Lambda (x) = 0.55\, x^2 + 0.16 \, x^3 + 0.29 \, x^6$ & 3.3 & $4.97$ \\
        \bottomrule
    \end{tabular}
\end{table}

\section{Conclusions}
\label{sec:Conclusions}
The increasing interest on grant-free protocols for \acl{MMA} in next generation wireless networks requires reliable design tools in order to optimize and compare different solutions. 
We propose a joint \ac{PHY} and \ac{MAC} layer design targeting a massive \ac{MIMO} system over Rayleigh fading channel.
The design is presented step-by-step to facilitate extending the analysis to other \ac{PHY} processing strategies and channel assumptions.
As a main outcome of our analysis we show that \ac{IRSA} distributions, optimum over the simple collision channel model, may turn suboptimum under a realistic setting. 



\section*{Acknowledgements}
This work has been carried out in the framework of the CNIT National Laboratory WiLab and the WiLab-Huawei Joint Innovation Center.

\begin{table}[t]
    \centering
    \caption{Optimum asymptotic thresholds constrained to $\Lambda'(1) = 3.0$ versus the number of antennas $M$. Optimum distribution $\Lambda(x) = x^3$ in all cases, $N_\mathrm{P} = 64$, $t = 10$, $N_\mathrm{D} = 256$.}
    \label{tab:VarM}
    \small
    \begin{tabular}{ccc}
        \toprule
        \rowcolor[gray]{.95}
        $M$ & $G^*$ & $G^* / M$\\
        \midrule
        $8$   & $0.1356$ & $0.0169$ \\
        $16$  & $0.4409$ & $0.0276$ \\
        $32$  & $1.0562$ & $0.0330$ \\
        $64$  & $2.0778$ & $0.0325$ \\
        $128$ & $3.8167$ & $0.0298$ \\
        $256$ & $6.9909$ & $0.0273$ \\
        \bottomrule
    \end{tabular}
\end{table}



\end{document}